\documentclass[aps,prb,reprint,amsmath,amssymb,superscriptaddress]{revtex4-2}

\usepackage{graphicx}
\usepackage{dcolumn}
\usepackage{hyperref}
\usepackage{bm}
\usepackage[utf8]{inputenc}
\usepackage[T1]{fontenc}
\usepackage{color}
\usepackage{rotating}

\begin{document}
\title{Specific heat studies of the phase transitions in multiferroic Sr$_{1-x}$Ba$_x$Mn$_{1-y}$Ti$_y$O$_3$ system ($0 \le x \le 0.65, 0 \le y \le 0.1$)}

\author{J.~Wieckowski}

\author{A.~Szewczyk}
\email{szewc@ifpan.edu.pl}

\author{M.~U.~Gutowska}

\author{B.~Dabrowski}

\affiliation{Institute of Physics, Polish Academy of Sciences, Aleja~Lotników 32/46, PL-02668 Warsaw, Poland}

\begin{abstract}
Specific heat studies of the Sr$_{1-x}$Ba$_x$Mn$_{1-y}$Ti$_y$O$_3$ polycrystalline samples performed by the relaxation and DSC methods over the temperature range 2 - 450 K are reported. Anomalies accompanying the antiferromagnetic-paramagnetic and ferroelectric-paraelectric phase transitions were measured and analyzed.  The system studied is a promising multiferroic material of the 2$^{\text {nd}}$ type, in which a strong coupling between the magnetic and electric systems can be related to the fact that the same Mn$^{4+}$ ions are responsible for the two orderings. Analysis of the anomaly at the magnetic transition was done by using the advanced theory of the continuous transitions, in which the presence of higher order terms in the free energy is considered, and the parameters of the critical behavior of the system were estimated. It was found that, despite of earlier predictions, the transition loses the continuous character and can become the 1$^{\text {st}}$ order process for the Ti-containing samples. The anomalies accompanying the 1$^{\text{st}}$ order high-temperature ferroelectric transition were found to be hardly visible by using both techniques, which was ascribed to a wide temperature range of coexistence of the dielectric and ferroelectric phases.
\end{abstract}

\maketitle

\section{Introduction}
By the definition \cite {schmid}, "multiferroics"  are the materials, which display several coexisting long range orderings, often of the magnetic and dielectric nature. Usually, one distinguishes the 1$^{\text{st}}$ type or "proper" multiferroics, in which the long range orderings appearing at separated temperatures are fairly independent of each other, and the 2$^{\text{nd}}$ type or "improper" ones, in which the orderings are strongly coupled or even the appearance of a one ordering is induced by the appearance of the second one, as for example in TbMnO$_3$ \cite {kenzelmann, mostovoy, kimura}. The latter multiferroics, especially magnetoelectrics, attract very large interest of researchers, because they could find numerous applications, including electrically controlled magnetic memories, electric amplifiers of magnons, magnetic field sensors, writing/reading heads of hard drives, and so on \cite{pyatakov}.

Presumably, the group of $AB$O$_3$ oxides having a crystalline structure of a slightly deformed perovskite (where $A$ = alkaline or rare earth ion or a set of such ions and $B$ = transition metal ion or a set of such ions), among which there are more than one hundred compositions showing magnetic orderings and a similar number of compositions showing a ferroelectric ordering, should be promising set for obtaining many 2$^{\text {nd}}$ type multiferroic compounds. However, there are only few known multiferroic compositions of this group. It is believed that the scarceness is due to the crystalline structure of these materials, in which each $B$ ion is located inside octahedral coordination made of 6 oxygen ions. In the ideal perovskite structure, the $B$-O$_6$  octahedra are relatively rigid and densely packed complexes forming a simple corner-shared cubic lattice, whereas the $A$ ions locate in free space between the octahedral complexes. In many materials, due to misfit of the atomic radii of the $A$, $B$ and O$^{2-}$ ions, small deformations and rotations of the $B$-O$_6$ complexes are observed, which lower crystal symmetry from the cubic $Pm\overline{3}m$ one of the ideal perovskite to a tetragonal, rhombohedral, or orthorhombic one, however, the general integrity of the perovskite-based network remains unchanged.

Calculations based on the density functional theory showed \cite{khomskii} that it is energetically favorable for the $B$ ions with the filled $d$ shell to shift from the center of the oxygen octahedron towards one of the apical oxygen ions. This breaks the inversion symmetry of the distribution of positive and negative ionic charges resulting in appearance of a nonzero spontaneous electric polarization and, in some cases, in the appearance of the ferroelectric ordering. On the contrary, if  $B$ is a "magnetic" ion, i.e. if it has a partially filled $d$ shell, the intraatomic exchange interactions (i.e., Hund's coupling) must be taken into account, which leads to the conclusion that the location in the center of the octahedron is the most favorable position for such an ion. As the result, the inversion symmetry is preserved and the material displays no spontaneous dielectric polarization. However, some kinds of a magnetic ordering can appear in such compounds.

In this context, researchers intensively seek for methods of overcoming this incapability and obtaining the multiferroic perovskites. For example, the following methods were described in the literature \cite{khomskii}:  (i) partial replacement of magnetic $B$ ions with nonmagnetic ones, which resulted in discovering the PbFe$_{0.5}$Nb$_{0.5}$O$_3$ multiferroic; (ii) separating magnetic properties with the $B$ sublattice and the ferroelectic properties with the $A$ sublattice occupied by ions having, like Bi, so called, lone pair of $s$ electrons, which is the origin of multiferroic properties of intensively studied BiFeO$_3$ and BiMnO$_3$ compounds, and (iii) synthesizing materials, like TbMnO$_3$, in which an incommensurate spiral magnetic structure appears for which, due to the Dzyaloshinski-Moriya interactions, displacement of magnetic and oxygen ions from their "normal" symmetric positions, producing nonzero electric polarization, accompanies the appearance of the magnetic spiral structure \cite {kenzelmann, mostovoy, kimura}. However, it was found that, with the exception of the (iii) group, in all such materials, the magnetic and electric order parameters are coupled rather weakly.

	The new method of obtaining multiferroic perovskites, proposed in Ref. \cite{sakai}, was based on the idea that: (i) much stronger magnetoelectric coupling could be achieved, if the same ions were responsible for the electric and magnetic orderings, and (ii) stretching the oxygen octahedra by the chemical tensile stress, produced by replacement of small A ions with larger ones could result in producing more free space around the magnetic ions, e.g., Mn$^{4+}$, located within the octahedra, and in making the displacement of the magnetic ions from the centers of the octahedra energetically favorable, which would lead to the appearance of the spontaneous polarization. This idea was verified for the Sr$_{1-x}$Ba$_x$MnO$_3$ system, \cite{sakai} and \cite{somaily}, by replacing the smaller Sr$^{2+}$ ions with the larger Ba$^{2+}$ ones. It was found, that indeed, for $x \geq 0.45$ the structural transition to the ferroelectric phase appears at high temperatures, $T_C \sim400$~K, and then, on lowering temperature, at $T_N \sim200$~K, the phase transition to the antiferromagnetic phase of the type G takes place and thus, the multiferroic state is gained. Unfortunately, it turned out that the system is not the ideal one, because it shows too high electric conductivity, which prohibits a direct measurement of the spontaneous polarization, and the structural deformation resulting in the spontaneous polarization decreases significantly at the point of magnetic transition. Nevertheless, this reduction of deformation proves the existence of a desired strong coupling betwee of the magnetic and electric orderings, which makes the Sr$_{1-x}$Ba$_x$MnO$_3$ system a good initial composition for further investigations. The continued studies of the Sr$_{1-x}$Ba$_x$MnO$_3$ system established the synthesis procedures allowing for increasing the limits of Ba substitution and the partial replacement of Mn with Ti, which resulted in stronger ferroelectric deformation and higher $T_C$ \cite{chapagain}. Detailed studies of homogeneity and magnetic properties of the obtained compositions were reported in Ref.~\cite{chapagain}. The recent studies of the local atomic and magnetic structures by using x-ray and neutron pair distribution function analysis, polarized neutron scattering, and and muon spin relaxation technique showed presence of the nanoscale tetragonal distortions in the paraelectric phase and the short-range antiferromagnetic correlations in the paramagnetic state over the wide temperature ranges \cite {jones}.

The present studies were aimed at a closer analysis of antiferromagnetic and ferroelectric phase transitions appearing in the (Sr,Ba)(Mn,Ti)O$_3$ compounds and of the critical behavior near the antiferromagnetic transition. Specifically, we wanted to verify the suggestion expressed in Ref. \cite {chapagain}, that the substitution of Mn with Ti does not change the second order character of the magnetic transition. Since it is known \cite {chapagain} that at the magnetic transition point, the structural deformation responsible for the spontaneous electric polarization decreases substantially, it is obvious, that this transition influences also the lengths and angles of the Mn-O-Mn bonds, and consequently, the strength of exchange interactions. As it was shown in Ref. \cite {bean}, such effect, when sufficiently strong, can change the order of the transition from the 2$^{\text{nd}}$ to the 1$^{\text {st}}$ one.

Since the magnetic ordering appearing at the low temperature transition to the multiferroic state has the antiferromagnetic character and the high temperature ferroelctric-paraelectric phase transition weakly influences magnetic properties of the system, magnetic measurements are not appropriate for studying these transitions. As an alternative, the measurements of specific heat are more suitable and informative. Moreover, the sign and value of the specific heat critical exponent $\alpha$ depend strongly on the character of magnetic structure of the system studied \cite {oleaga}. Thus, the specific heat measurements were chosen as the main experimental method of the present studies and the detailed measurements of this parameter for a series of (Sr,Ba)(Mn,Ti)O$_3$ compositions, as well as the detailed analysis of the shape of the specific heat anomalies accompanying the antiferromagnetic phase transitions were performed.

\section{Experiment}
The almost single-phase polycrystalline Sr$_{1-x}$Ba$_x$Mn$_{1-y}$Ti$_y$O$_3$ samples with $x$ = 0 - 0.65 and y = 0.0 - 0.1, were synthesized from stoichiometric mixtures of BaCO$_3$ (99.9\% pure), SrCO$_3$ (99.99\%), MnO$_2$ (99.95\%), and TiO$_2$ (99.99\%) by methods described in detail in Refs. \cite{somaily} and \cite {chapagain}. In short, the synthesis of the perovskite compounds studied here is a difficult task involving repeated firing under increasingly reducing conditions at high temperatures to achieve the required chemical compositions surviving beyond ordinary solubility limits by removing the hexagonal second phases of the similar compositions, which normally form in the oxygen containing atmospheres. The procedure is followed by the oxidation of the oxygen deficient perovskites to the full oxygen content at lower temperatures without decomposition, however it could lead in some cases to the presence of small amounts of second phases and the chemical compositional inhomogeneities.

For the present specific heat measurements, the following compositions were chosen: $y$ = 0 and $x$ = 0, 0.43, and 0.45; $y$ = 0.02 and $x$ = 0.5; $y$ = 0.06 and $x$ = 0.45, 0.5, 0.55, 0.6 and 0.65; and $y$ = 0.1 and $x$ =  0.5.

Two methods of specific heat measurement were used. For the temperature range $2 \leq T \leq 400$~K the studies were done by means of the relaxation method by using the specific heat, HC, option of the PPMS system (made by Quantum Design). The measurements were performed for zero and several fixed values of the external magnetic field, $B$, up to 9 T. Typically, in the vicinity of phase transitions and below 15 K, the experimental points were measured every 0.2 or 0.5~K, whereas outside these regions, every 1, 2 or 3~K. To maintain the readability of all figures presented below, only the limited numbers of experimental points were denoted with the symbols. The high accuracy (to 2\%) of the determined absolute values of the specific heat for $T$ smaller than $\sim250$~K is the main advantage of this method. Above $\sim250$~K the heat transfer between the calorimeter and the environment via radiation increases substantially (it is proportional to $T^4$) and dominates the other mechanisms (convection and conduction), which results in increased noise and deterioration of the accuracy of the determined absolute specific heat values to ca. 5\%. Moreover, the inherent property of the relaxation method, i.e. the necessity of applying the measuring heat pulses increasing the sample temperature by ca 2\% of $T$   results in rounding the specific heat maxima accompanying phase transitions and makes this technique not optimally suited for determining the critical exponent $\alpha$.

Thus, for measurements in the temperature range $100 < T < 450$~K, the differential scanning calorimetry system, DSC (made by TA Instruments) was used. The DSC method, in which changes of heat capacity are measured during the temperature sweep with a fixed speed (usually from 1 to 10~K/min) is very precise and effective in detecting all anomalies of the specific heat, usually those related to phase transitions. However, it does not allow to determine accurately the absolute specific heat values.

In this context, by combining the two methods mentioned above we were able to study both the magnetic phase transition appearing in the range $130 < T < 230$~K and the ferroelectric transition appearing in the range $\sim350 < T <\;\sim400$~K. Furthermore, by using the fact that the temperature ranges of applicability of the two measuring systems overlapped within the range $\sim100 < T <\;\sim350$~K we were able to shift the absolute specific heat values reported by the DSC system to the values determined by the relaxation method and thus to determine more reliable the absolute specific heat values for the high-temperature range $T > 300$~K.
	
Additionally, magnetization measurements were performed to verify the accuracy of the measured transition temperatures by using the vibrating sample magnetometer, VSM, option of the PPMS system.

\section{Results}
\begin{figure}
\includegraphics[width=0.5\textwidth]{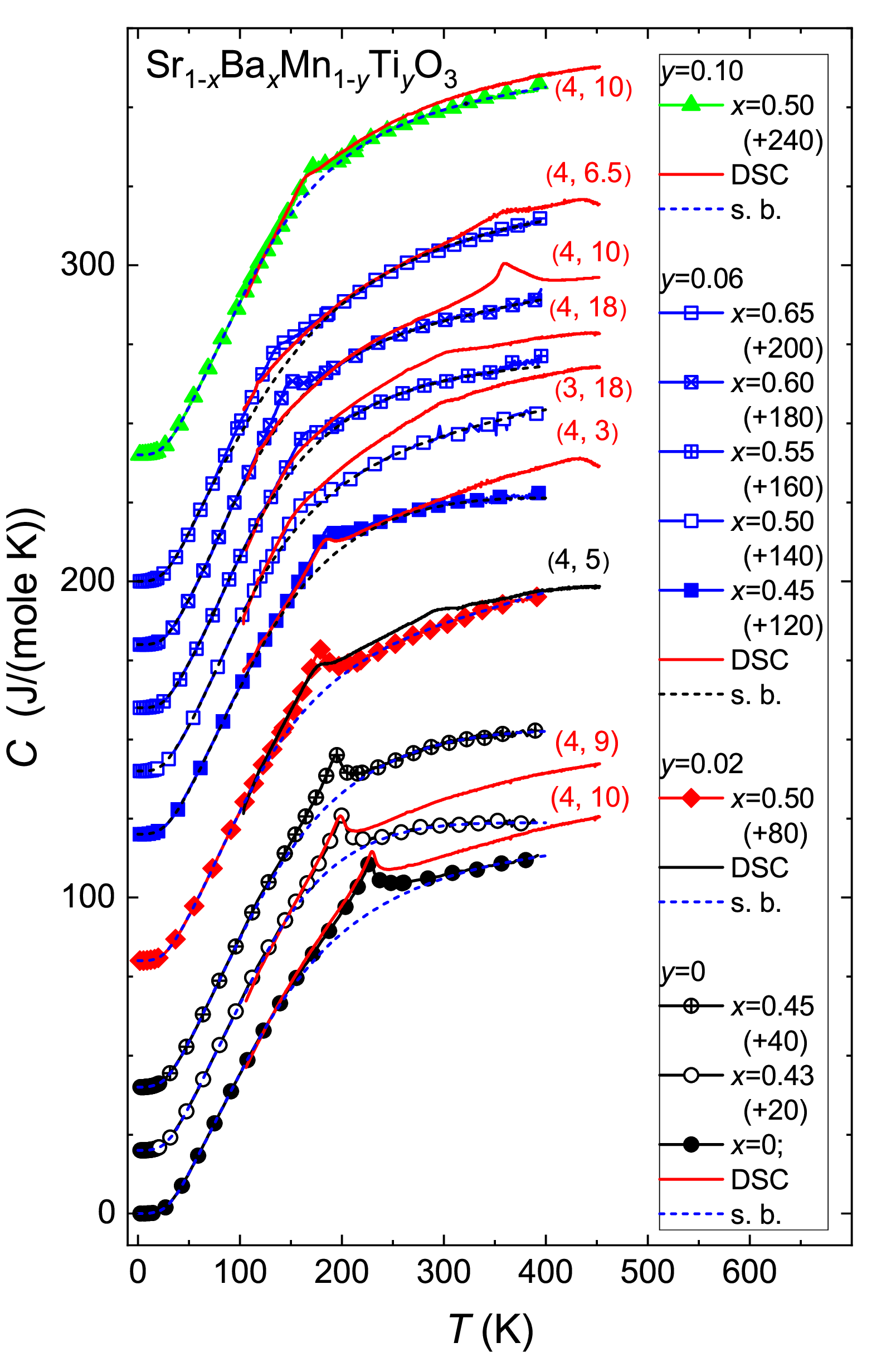}
\caption{\label{fig01}\  Specific heat of the Sr$_{1-x}$Ba$_x$Mn$_{1-y}$Ti$_y$O$_3$ compositions as a function of temperature. Curves measured with the PPMS are denoted by symbols and plotted in black, red, blue, and green color, for $y$=0, 0.02, 0.06, and 0.10, respectively. On each PPMS curve, there are superimposed: (i) the smooth background, s.b., mimicked with Eq. \eqref{eq1}, plotted with the dashed line (black for $y$ = 0.06 and blue for other $y$ values), and (ii) the dependence measured with the DSC method, plotted with the solid line (black for $y$ = 0.02 and red for other $y$ values). In the legend, at the symbols denoting the PPMS data, the $x$ content and next, in parentheses, the value by which all curves for this composition are shifted along the vertical axis are given. At each DSC curve the heating rate at which it was registered, in K/min, and next, the value in J/(mole K) by which the DSC curve was shifted with respect to the PPMS data to get the more reliable absolute specific heat values are given in parentheses.
}
\end{figure}

\begin{figure}
\includegraphics[width=0.5\textwidth]{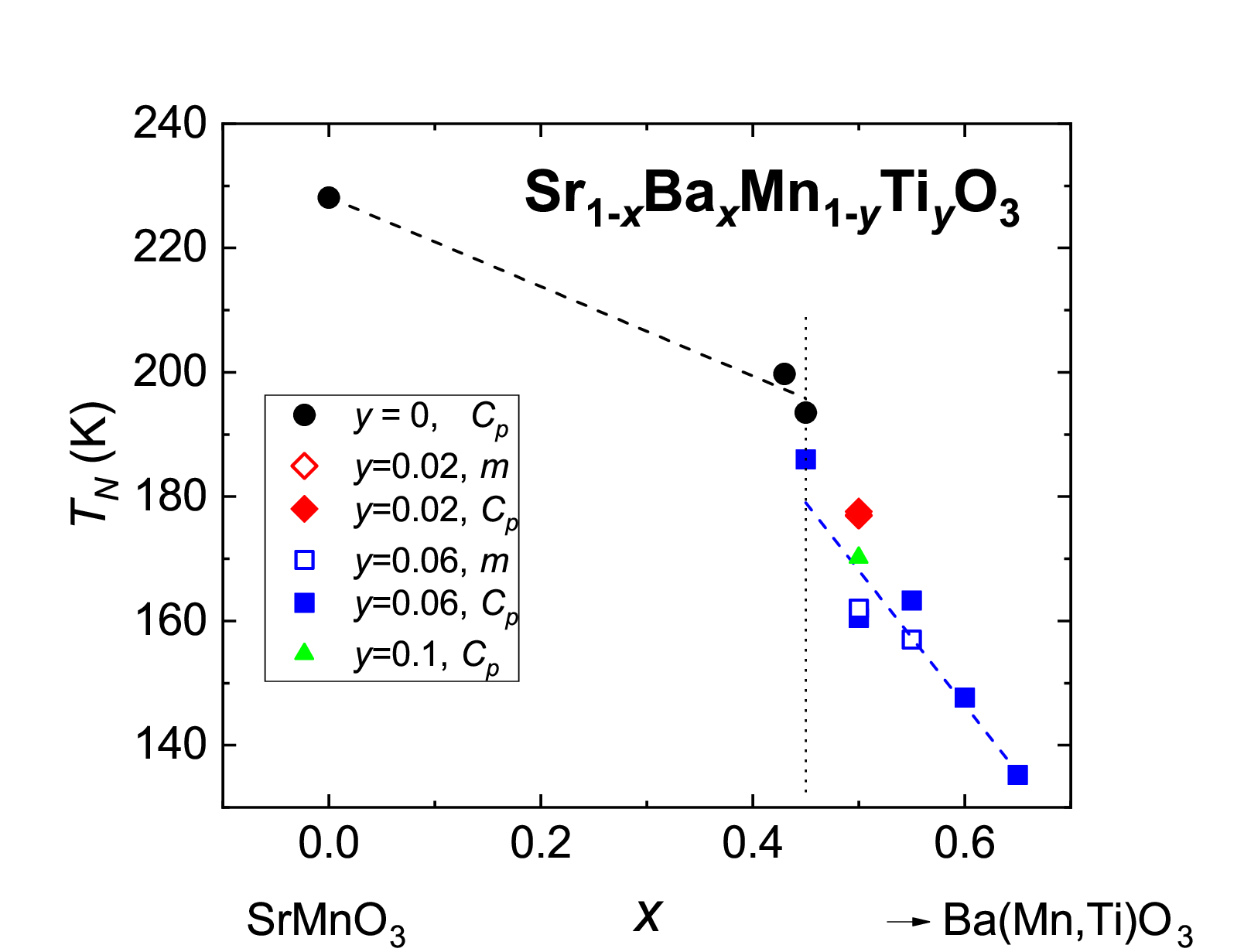}
\caption{\label{fig02}\ The Néel temperatures of the Sr$_{1-x}$Ba$_x$Mn$_{1-y}$Ti$_y$O$_3$ system, $T_N$, as a function of $x$. The arrow indicates a hypothetical composition to which the system evolves with increasing $x$.
}
\end{figure}

\begin{figure}
\includegraphics[width=0.5\textwidth]{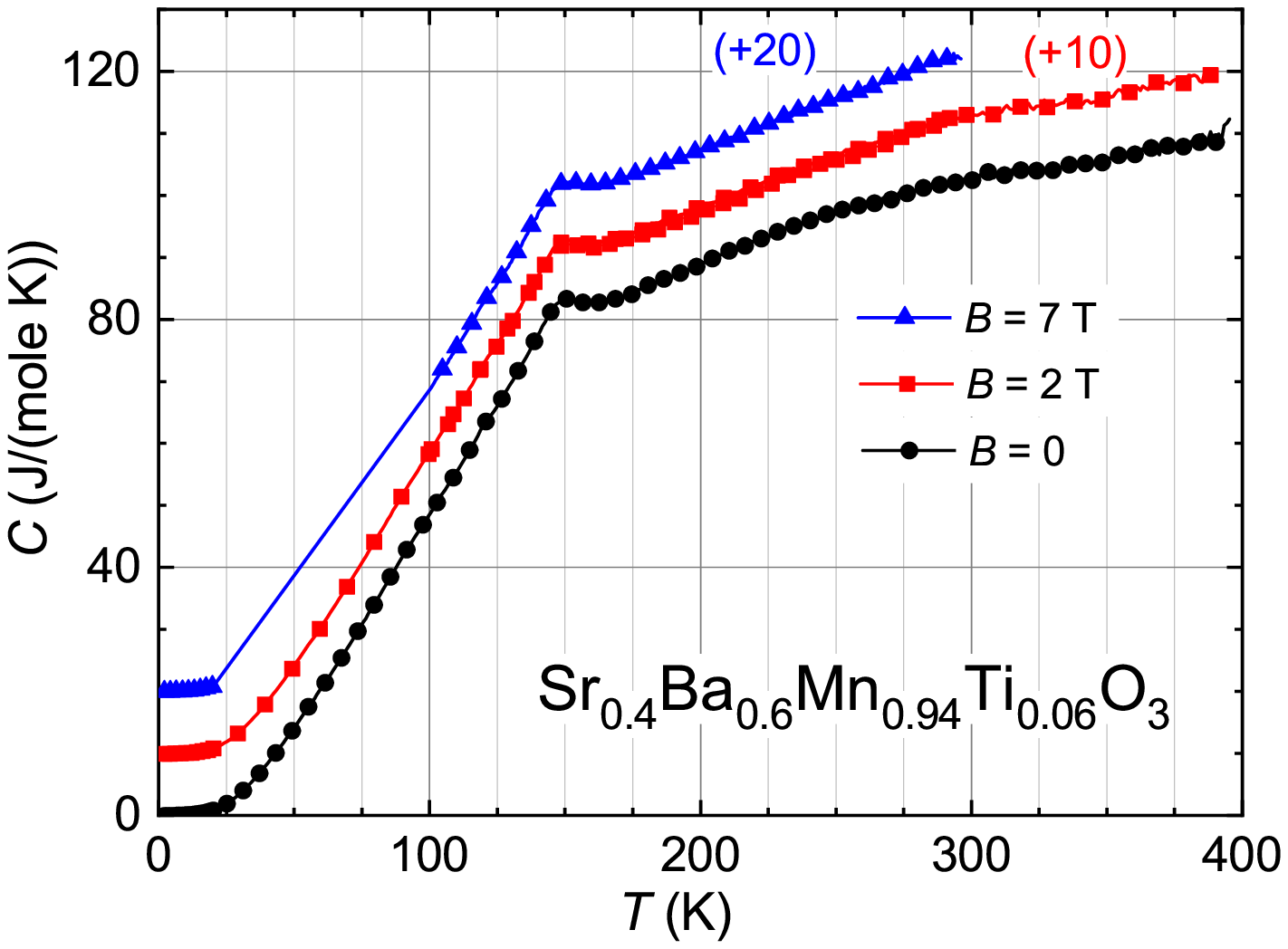}
\caption{\label{fig03}\ Temperature dependence of the specific heat, $C$, of Sr$_{0.4}$Ba$_{0.6}$Mn$_{0.94}$Ti$_{0.06}$O$_3$ in the presence of an external magnetic field, $B$. To maintain legibility, only selected measured points are indicated by symbols and the curves are shifted along the vertical axis by the values given in parentheses.
}
\end{figure}

Specific heat, $C$, measured as a function of temperature in zero magnetic field for all compositions studied is presented in Fig.~\ref{fig01}. One can easily notice that the most pronounced anomaly accompanies the phase transition, which was identified earlier \cite{sakai, somaily, chapagain} as the transition from the paramagnetic to the antiferromagnetic phase of the type G. It was found that replacement of both Sr with Ba and Mn with Ti lowers the N\'{e}el temperature, $T_N$, from 228~K for the pure SrMnO$_3$ to 135~K for the Sr$_{0.35}$Ba$_{0.65}$Mn$_{0.94}$Ti$_{0.06}$O$_3$ composition, see Fig.~\ref{fig02} and Table~\ref{tab2}. In all the cases, the specific heat anomaly is wide, smeared over the temperature range of 110-220 K depending on the composition. We attribute the smearing to small inhomogeneities of the chemical composition of particular samples and to small deviations of the actual oxygen stoichiometry from the value "3", which could appear due to, mentioned above, technological difficulties in synthesizing these compositions.

Usually, in the case of antiferromagnetic transitions, the uniform external magnetic field lowers the transition temperature and sharpens the specific heat anomaly that it accompanies. However, the specific heat measurements performed for the (Sr,Ba)(Mn,Ti)O$_3$ system for several fixed values of the magnetic field, $B$, up to 9~T showed that the influence of the magnetic field on the antiferromagnetic transition is always negligible, even for $B =9$~T. As an example, this effect is illustrated in Fig.~\ref{fig03}, where the specific heat of the $x$=0.6 and $y$=0.06 composition, measured as a function of temperature for several fixed $B$ values is presented. This is rather a surprising effect, because the molecular field for the studied system, roughly estimated based on the $T_N$ value, is of the order of 35 - 60 T and the external field of 9 T constitutes its noticeable fraction. We presume that the influence of $B$ on the transition seems to be negligible because of the large widths of the anomalies related to the antiferromagnetic transition.

Unlike the antiferromagnetic transitions, the specific heat anomalies at the ferroelectric transitions, are, practically, invisible in the PPMS data and hardly visible in the DSC results, Fig.~\ref{fig01}. We attribute this to the fact that these transitions are of the first order and, as the x-ray studies showed \cite{chapagain}, the temperature range of coexistence of para- and ferroelectric phases is relatively wide ($\sim60$~K), which smears the specific heat anomaly related to the ferroelectric transition.

\section{Analysis}
To investigate the anomalies accompanying the antiferromagnetic phase transition it was necessary to extract the largest contribution to the measured specific heat, i.e., the noncritical smooth background, composed of the phonon and magnon contributions. Since in the case of weekly anisotropic antiferromagnetic materials, dispersion relation for the magnons has same functional form as for the phonons ($\omega \sim k$), the magnon and phonon contributions to specific heat depend on temperature in the same way and separating them is not trivial. Thus, to mimic the effective smooth contribution to the specific heat we used the following, frequently applied expression \cite{zajarniuk, wiec}, consisting of a mixture of the Debye and Einstein models:
\begin{equation}
\begin{split}
&C_s\left (T\right )= \Bigg [ \frac{3 n_{D} k_B N_ A}{(1 - \alpha_D T)}  \left ( \frac{T}{\theta_{D}}\right )^3
\int_{0}^{{\theta_D}/{T}} \frac{x^4e^x}{(e^x - 1)^2} dx \\
&+ \frac{k_B N_ A}{(1 - \alpha_E T)} \sum_{i = 1}^{n_{O}} n_{i} \left ( \frac{\theta_{ i}}{T} \right ) ^{2} \frac{e^{{\theta_i}/{T}}}{(e^{{\theta_{i}}/{T}} - 1)^2} \Bigg ].
\label{eq1}
\end{split}
\end{equation}
\noindent The first term of \eqref{eq1} describes the contribution arising from the $n_D$ branches approximated by the Debye model with the Debye temperature $\theta_D$ and the second term represents the contribution arising from the $n_O$ optical branches described by the Einstein model. The Einstein temperature $\theta_i$ and the number $n_i$ of modes are ascribed to each of the $n_O$ optical branches. The effect of lattice expansion can be considered in the way proposed in Ref.~\cite {martin}, i.e, by introducing the factor   $1/ (1- \alpha_i T)$ for each branch. In the present analysis we introduced a simplification, consisting in considering only two $\alpha_i$ coefficients, namely, the one for the Debye branches, $\alpha_D$, and the other one, $\alpha_E$, for the Einstein branches. Satisfactory approximation of the smooth specific heat background for each composition has been obtained by taking $n_D=3$ and $n_O=2$,    and treating $\theta_D$, $n_i$, $\theta_i$, $\alpha_D$, and $\alpha_E$ as the fitted parameters. The modeled smooth backgrounds for all the compositions studied are plotted in Fig.~\ref{fig01} with the dashed lines. The values of fitted parameters for which the best descriptions of the backgrounds were achieved are collected in Table~\ref{tab1}. These values have only the meaning as mathematical fitting parameters, so their uncertainty was not estimated.
\begin{table}
\footnotesize{
\caption{\label{tab1} Fitting parameters used for modeling the smooth specific heat background of Sr$_{1-x}$Ba$_x$Mn$_{1-y}$Ti$_y$O$_3$ compositions.}
\begin{ruledtabular}
\begin{tabular}{cc|ccccccc}
\multicolumn{2}{c|}{sample}
 & $\theta_D$ & $\alpha_D$ & $\alpha_E$ & $\theta_1$ & $n_1$ & $\theta_2$ & $n_2$ \\
\ $x$ & $y$ & (K) & (1/K) & (1/K) & (K) & & (K) & \\
% & & & & & & & & \\
\hline
% & & & & & & & & \\
 0.5 & 0.1 & 231.7 & 9.8 $\cdot10^{-4}$ & -9.2$\cdot10^{-4}$ & 256 & 3 & 514 & 11 \\
\hline
 0.65 & 0.06 & 223 &  -2.8$\cdot10^{-3}$ & 1.6$\cdot10^{-4}$ & 268 & 5 & 581 & 8 \\
 0.6  & 0.06 & 294.6 & 1.05$\cdot10^{-3}$ & -8.79$\cdot10^{-4}$ & 177.7	& 2 & 447.3	& 10 \\
 0.55  & 0.06 & 247 & -11.4$\cdot10^{-3}$ & -8$\cdot10^{-5}$ & 205.2 & 3 &	446.4 &	11 \\
 0.5  &0.06  & 253	& -4.15$\cdot10^{-3}$ &	3.14$\cdot10^{-4}$ & 191 & 2 & 430 & 10 \\
 0.45 & 0.06 & 230.7 & -4$\cdot10^{-3}$ & -5.5$\cdot10^{-4}$ & 277 & 5 & 550 & 11 \\
\hline
 0.5 & 0.02 & 260 & 1.14$\cdot10^{-3}$ & -6.5$\cdot10^{-4}$ & 208.9 & 2 & 474.3 & 10 \\
\hline
 0.45 & 0  & 242 & -3.9$\cdot10^{-4}$ & -5.16$\cdot10^{-4}$ & 265 & 4 &	554	& 11 \\
 0.43 & 0 & 260 & 9$\cdot10^{-4}$ & -5.06$\cdot10^{-3}$ & 267.3 & 5 & 612 &	23 \\
 0 & 0 & 267 & -4.8$\cdot10^{-3}$ & 1.9$\cdot10^{-5}$ & 246.9 & 4 & 534.3 & 10 \\
\end{tabular}
\end{ruledtabular}
}
\end{table}

\begin{figure*}
\includegraphics[width=1\textwidth]{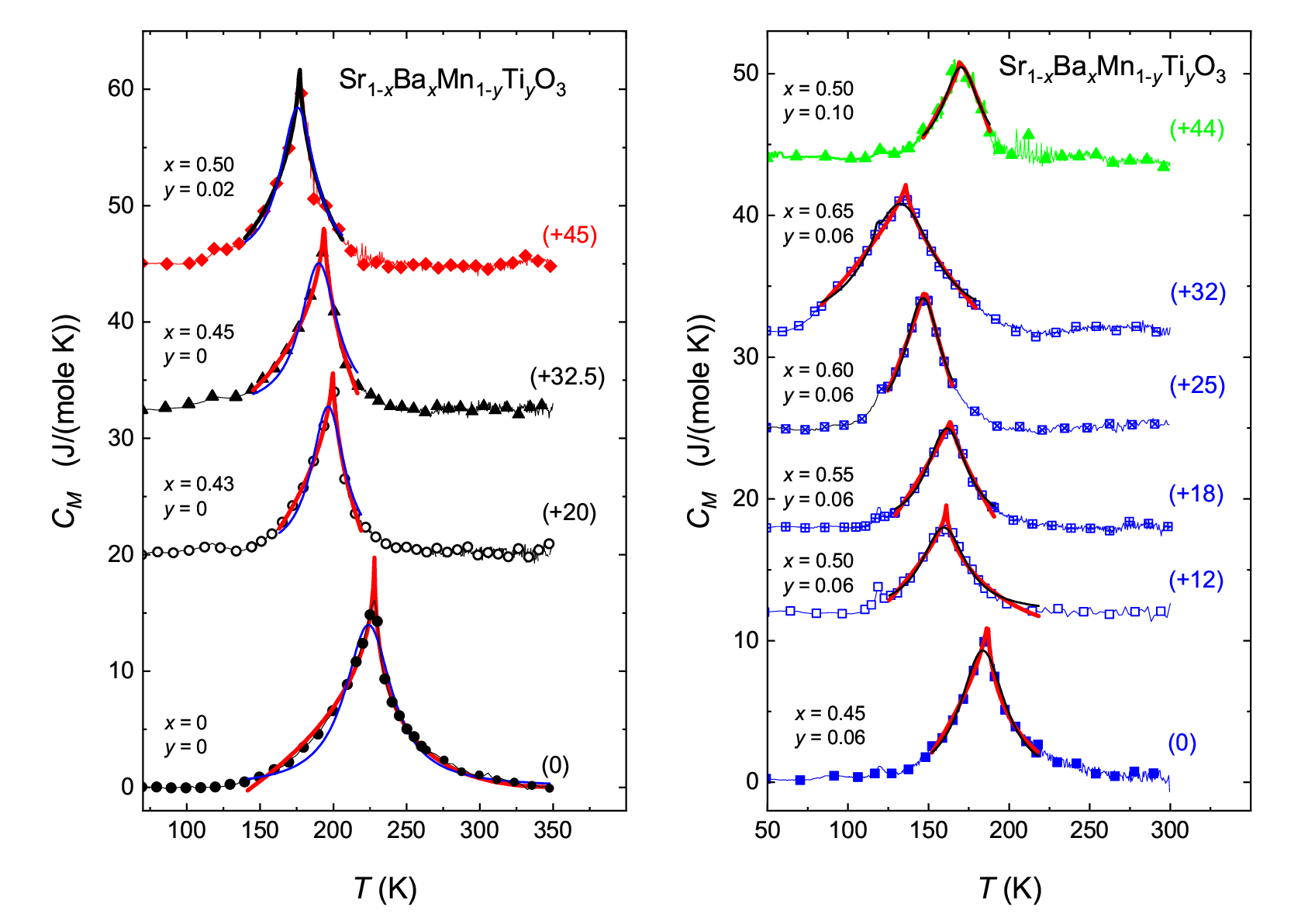}
\caption{\label{fig04}\  Specific heat anomalies related to the antiferromagnetic phase transition appearing in the Sr$_{1-x}$Ba$_x$Mn$_{1-y}$Ti$_y$O$_3$ compositions. On each experimental dependence, denoted with the same symbols as in Fig.~\ref{fig01}, the fitted theoretical critical dependence \eqref{eq2}, taking into account higher order terms, plotted with the thick solid line (blue for $y$=0.02 and red for all the other $y$ values), as well as the fitted Lorentz function, plotted with the thin solid line (blue for $y$=0 and black for all the other $y$ values) are superimposed. Above each experimental curve, the $x$ and $y$ values, as well as the values, in parentheses, by which this curve was shifted along the vertical axis to maintain readibility, are given.
}
\end{figure*}

The specific heat anomalies, $C_M(T)$, related to the magnetic phase transition were derived by subtracting the modeled smooth backgrounds from the measured temperature dependences of the specific heat. They are presented in Fig.~\ref{fig04}. As mentioned above, these anomalies are smeared over the wide temperature ranges and thus, they could not be analyzed in frames of the basic theory of critical phenomena, in which one assumes that the critical behavior given by a single power function appears when the parameter    $\tau= \left ( T - T_N \right ) / T_N$ takes the values $|\tau|\leq 10^{-3}$. Apart from the wide temperature spread of the anomalies, analysis of their real shape within the critical range $|\tau|\leq 10^{-3}$ was, in practice, impossible, because they were measured by means of the relaxation method. As discussed previously, in this method, during the measurement process a measuring heating pulse, increasing the sample temperature by the value of the order of 2\% of the actual $T$ value is applied. Thus, the specific heat value assigned to the $T+(2\%\cdot T/2)$ temperature is, in fact, the value averaged over the temperature range from $T$ to $T + 2\% \cdot T$, which results in "rounding" the anomalies measured for $|\tau|\leq 10^{-3}$.

Therefore, following Refs. \cite {oleaga, wegner, ahlers1}, it was assumed that for the compositions studied, the more advanced theory of critical phenomena, taking into account the presence of higher order terms in $\tau$ in the expression for the free energy, essential for $|\tau|\geq 10^{-3}$, must be applied. According to this theory, the critical behavior of the specific heat can be represented in the form:
\begin{equation}
C_M\left (T\right )=\frac {B} {\alpha} \left [ \left | \frac {T} {T_N} -1 \right |^{-\alpha} -1 \right ] \left [ 1 + E \left | \frac {T} {T_N} -1 \right | ^x \right ] + D,
\label{eq2}
\end{equation}

\noindent which differs from the classical one by the presence of the term in the second parentheses.

In fact, some authors, e.g. \cite{ahlers1, kornblit1}, instead of modeling the smooth regular background, e.g. by using Eq.~\eqref{eq1}, prefer to take it into account by adding to Eq.~\eqref{eq2} a linear term of the form: $D_1+F\cdot T$, with $D_1$ and $F$ being constant values. However, we believe that for the anomalies smeared over such large temperature range as in the present case, modeling the smooth background by \eqref{eq1} is more appropriate than a linear approximation.

We verified that for the system studied, the best description of the experimental anomalies with \eqref{eq2} is achieved for the exponent  $x = 0.44$, and this value was used in the further analysis. This result is consistent with the earlier papers, e.g. \cite{kornblit2}, presenting similar analysis, which showed that for all real physical systems studied until now, the best description can be achieved by taking the value $x\simeq 0.5 \pm 0.2$.

The preliminary calculations showed that for describing the maximum appearing at the N\'{e}el temperature for all the compositions, the negative value of the exponent $\alpha$ must be chosen, which is reasonable, because negative $\alpha$ values are predicted also in some theoretical models. For example, the $\alpha = -0.08 \pm 0.04$ value is predicted theoretically for the Heisenberg model \cite{binney}. It is important to note that for the negative $\alpha$ values, specific heat does not diverge but shows a finite maximum at the transition point and hence, the $C_M(T)$ function must be continuous at $T_N$. Thus, if we denote the parameters of the function (\ref{eq2}) with the superscript "-" for $\tau <0$, i.e. for $T < T_N$, and with the superscript "+" for $\tau > 0 $, i.e. for $T > T_N$, the continuity requirement leads to the relation:
\begin{equation}
D^+ = D^{-} + \frac {B^{+} - B^{-}} {\alpha}.
\label{eq3}
\end{equation}
Next, we treated $\alpha, B^-, \:B^+, \:E^-, \:E^+, \:D^-$, and $T_N$ as fitted parameters and matched the dependence \eqref{eq2} to the experimental data for a rather wide temperature range around the phase transition, chosen judiciously for each composition. During the analysis of the main phase, when choosing the temperature range for fitting, we considered the width of the anomaly and the necessity of minimizing the influence of a small amount of a parasitic phase, visible near 120 K as a small maximum
on the low temperature side of the specific heat anomaly for majority of the compositions. As it is shown in Fig.~\ref{fig04} with the thick solid lines, the shape of the specific heat anomaly related to the magnetic phase transition can be reproduced very well with Eq.~\eqref{eq2} for all the compositions.

The values of the fitted parameters giving the best description of the experimental data are given in Table~\ref{tab2}. Uncertainty of each of these parameters was estimated by keeping all other parameters fixed and varying the investigated parameter up to the value for which a noticeable change of the fitted curve was observed. For completeness, the $D^+$ parameter, not fitted but calculated according to Eq.~\eqref{eq3}, is also given in the table. To differentiate the fitted $T_N$ value from the value determined experimentally, given in the last column of the table, the superscript "cr" was added to the former. Since in this analysis, the seven parameters were fitted, the $\alpha$ values given in Table~\ref{tab2} cannot be treated as the real, accurately determined values of the critical exponent. In fact, the performed analysis allows only to draw the conclusion that the $\alpha$ parameter is negative for the considered system.

%\begin{sidewaystable}
\begin{table*}
%\footnotesize{
\caption{\label{tab2} Parameters of the theoretical function \eqref{eq2} and the Lorentz function \eqref{eq4}, for which the best fit to the experimental specific heat anomalies related to the antiferromagnetic transition for the Sr$_{1-x}$Ba$_x$Mn$_{1-y}$Ti$_y$O$_3$ compositions was achieved.}
\begin{ruledtabular}
\begin{tabular}{cc|cccccccc|ccc|c}
\multicolumn{2}{c|}{sample}
 & $\alpha$ & $B^{-}$ & $B^{+}$ & $E^{-}$ & $E^{+}$ & $D^{-}$ & $D^{+}$ & $T_N^{cr}$ & $L_0$ & $w$ & $T_N^L $ & $T_N$ \\
\ $x$ & $y$ &  & \multicolumn{2} {c} {$( \text {J/(mole K))}$ } & & &\multicolumn{2} {c} {$(\text {J/(mole K))}$ } & $(\text {K})$ & $ \left ( \frac {\text {J}} {\text {mole K}} \right )$  & $(\text {K})$ & $(\text {K})$ & $(\text {K})$ \\
& & & & & & & & & & & & & \\
\hline
 & & & & & & & & & & & & & \\
 0.5 & 0.1 & -0.311(1) & 9.68(2) & 17(1) & 1.64(2) & 2.07(2)& -22.87(2) & -46(3) & 168.9(2)  & 6.5(1) & 28(1) & 170.2(5) &  170.2(5) \\
\hline
 0.65 & 0.06 & -0.76(1) & 13.4(1) & 13.4(2) & -0.003(3) & -0.22(4) & -7.36(6) & -7.4(5) & 135.9(5)  & 8.80(5) & 50.8(1) & 132.5(1) & 135.2(5)\\
   &   &   &   &   &   &   &   &   &    & 0.8(1)  & 3.7(4) & 118.1(1) & \\
 0.6  & 0.06 & -0.574(1) & 21.07(2) & 26.1(5) & 0.54(1) & 0.53(1) & -27.24(2) & -36.0(9) & 147.7(2) & 9.17(5) & 27.8(8) & 146.6(5) & 147.7(5) \\
 0.55  & 0.06 & -0.645(3) & 16.43(5) & 17.4(4) & 0.35(1) & 0.22(2) & -18.12(6) & -19.7(8) & 163.7(5)  & 6.97(5) & 30.7(7) & 161.4(4) & 163.3(5) \\
 0.5  &0.06  & -1.25(1) & 24.2(2) & 9.1(2) & -0.49(2) & -1.82(3) & -11.4(2) & 0.68(62) & 161.0(8) & 6.0(1) & 33.6(9) & 159.4(9) & 160.5(5)  \\
 0.45 & 0.06 & -0.80(1) & 21.9(2) & 5.4(1) & -0.27(2) & -3.48(5) & -15.5(2) & 5.1(9) & 186.8(5) & 9.3(2) & 34(2) & 184(1) & 186.0(5) \\
\hline
 0.5 & 0.02 & -0.90(5) & 11.3(5) & 11.9(2) & -2.92(5) & -3.08(5) & 5.4(2) & 4.7(9) & 177.0(5) & 13.4(2)& 29(2) & 175.7(6) & 177.1(5) \\
\hline
 0.45 & 0  & -0.76(3) & 15.7(3) & 25.4(6) & -1.14(3) & -0.87(3) & -3.5(2) & -16(2) & 193.7(5) & 12.6(5) & 31(2) & 190(1) & 193.5(5) \\
 0.43 & 0 & -0.66(1) & 19.9(2) & 26.0(8) & -0.48(4) & -0.56(3) & -13.4(2) & -23(2) & 199.6(5) & 12.7(5) & 28(2) & 196(1) & 199.7(5)  \\
 0 & 0 & -0.305(5) & 9.83(7) & 6.39(5) & 0.23(5) & -1.86(4) & -9.7(1) & 1.6(7) & 228.0(5) & 14.0(5) & 39(2) & 224(1) & 228.1(5)\\
\end{tabular}
\end{ruledtabular}
%}
\end{table*}
%\end{sidewaystable}

Fig.~\ref{fig04} shows that for the Ti-free compositions ($y=0$) the anomaly related to the antiferromagnetic transition is asymmetric and $\lambda$-shaped, which is typical of the 2$^{\text{nd}}$ order transitions, whereas with partial replacement of the Mn ions with Ti ions, the anomaly becomes more symmetric as is observed in the case of the 1$^{\text {st}}$ order transitions. To demonstrate this effect, we fitted  the Lorentz function:
\begin{equation}
L(T) = \frac {L_{0} w^2} {4 \left ( T - T_N^L \right )^2 +w^2 },
\label{eq4}
\end{equation}
to the experimental data for the same temperature ranges as that used for fitting by Eq.~\eqref{eq2}. The fitted Lorentz functions are plotted in Fig~\ref{fig04} by using thin solid lines and the fitting parameters: $L_0$, $w$, and $T_N^L$ are given in Table~\ref{tab2}. Like in the case of matching with Eq.~\eqref{eq2}, the uncertainty of each of the fitted parameters was estimated by keeping all other parameters fixed and varying the investigated parameter up to the value for which a noticeable change of the fitted curve was observed. Since for the sample with composition $x = 0.65, y =0.06$ the additional maximum related to small content of an unknown parasitic phase was located very close to the main maximum on its low temperature side, this composition was treated in a different way, i.e., the experimental maximum was described as a sum of two Lorentz functions, the first, which approximated the main anomaly, and the second, which estimated the parasitic anomaly. This approach reproduced quite well the shape of the combined anomaly and the fitted parameters of the two Lorentz functions are given in Table~\ref{tab2}.

\begin{figure}
\includegraphics[width=0.5\textwidth]{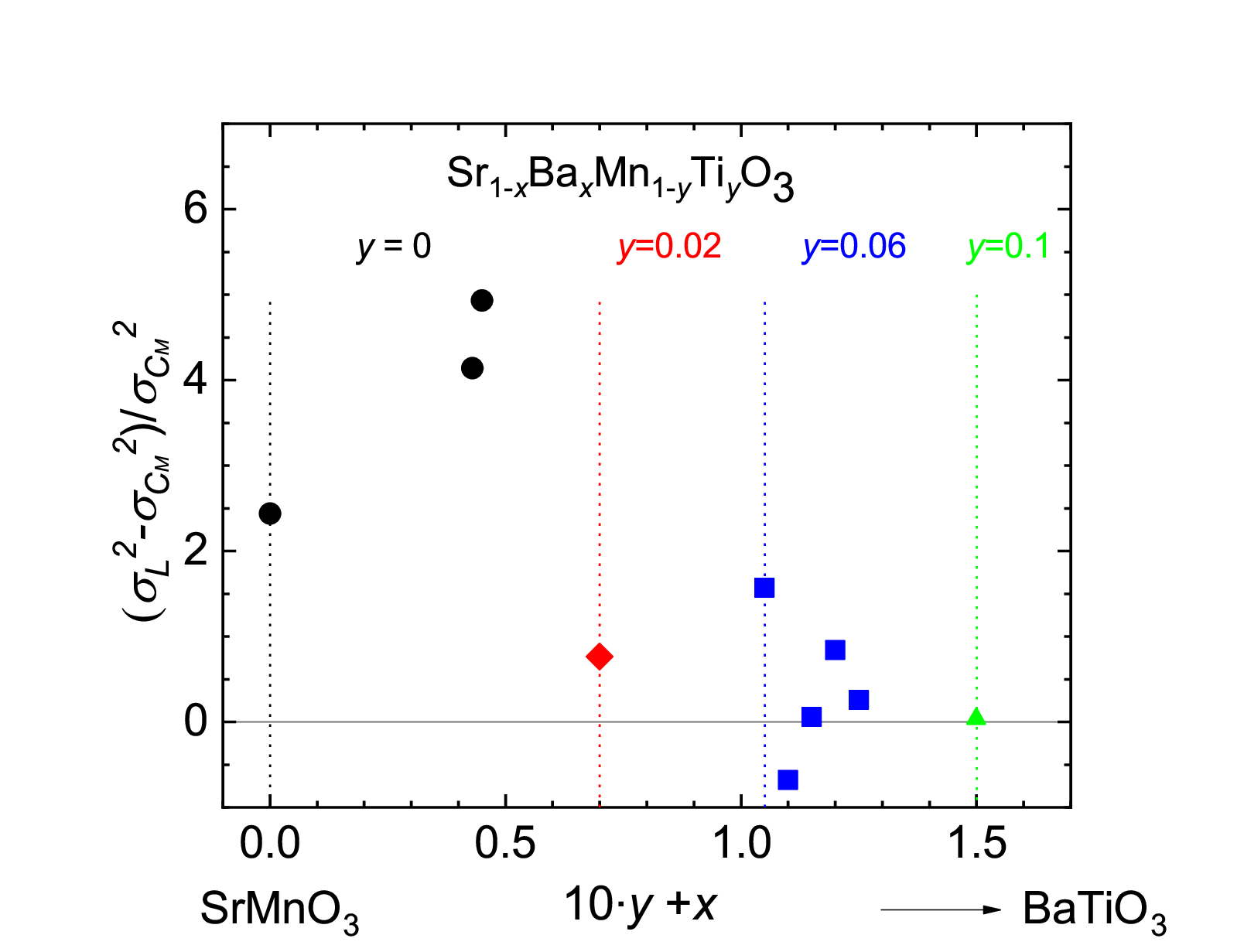}
\caption{\label{fig05}\ Difference between average square of residuals corresponding to the Lorentz function and the critical behavior of Eq.~\eqref{eq2} fittings as a function of composition. The arrow indicates the hypothetical composition to which the compositions studied evolve.
}
\end{figure}

We noticed that for many Ti-containing compositions, the Lorentz function describes the specific heat anomaly nearly as well as the curve calculated with Eq.~\eqref{eq2}. To evaluate this observation quantitatively, we considered an average square of residuals per one point, defined by the formula:
\begin{equation}
\sigma_f^2 = \frac {1} {N} \sum_{i=1}^N {\left( y_i^{ex} - f(x_i^{ex})\right )^2},
\label{eq5}
\end{equation}
where $f$ denotes the investigated theoretical function and the summation extends over all $N$ experimental points $(x_i^{ex}, y_i^{ex})$, for which the fitting was performed. This value has been calculated by taking as $f$ the $C_M(T)$ function given by \eqref{eq2} and by taking as $f$ the Lorentz function \eqref{eq4}. Next, we plotted the difference between  $\sigma_L^2$ and  $\sigma_{C_M}^2$ normalized to $\sigma_{C_M}^2$ as a function of the composition, identified by the parameter $10\cdot y + x$, in Fig.~\ref{fig05}.
Clearly, with increase in the Ti content the specific heat anomaly becomes more symmetric as it is expected for the case of a 1$^{\text {st}}$ order transition. In fact, with exception of the composition $x=0.5, y=0.06$, for which the Lorentz function describes the anomaly better, for all other compositions the description by Eq.~\eqref{eq2} is better. However, in our opinion, it is not related to the real physical effect but to the fact that the fits of Eq.~\eqref{eq2} were done with the larger number of free parameters. In conclusion, in our opinion, in spite of suggestions formulated in Refs.~\cite{chapagain, jones}, the antiferromagnetic phase transition in the (Sr,Ba)(Mn,Ti)O$_3$ system changes its character from the continuous (second order) to the discontinuous (1$^{\text {st}}$ order) transition with increase of the Ti content. This effect is very probably related to the complex nature of the antiferromagnetic magnetic transition, which is coupled to changes of the ferroelectric distortions.

\section{Conclusions}
Our specific heat studies of the (Sr,Ba)(Mn,Ti)O$_3$ compounds showed that the anomalies accompanying the antiferromagnetic transition are spread over the wide temperature range of  160~$\pm$~50~K. Thus, under the assumption that this is the $2^{\text {nd}}$ order transition, the critical behavior of the specific heat must be analyzed in frames of the more advanced theory than usually, i.e., within theory taking into account higher order terms in expansion of the free energy. Parameters of such critical behavior were determined. However, it was observed that with the increase of Ti content, the anomalies change their $\lambda$-shape by becoming more symmetric and can be approximated by the Lorentz function, which suggests that, in spite of the earlier predictions \cite{chapagain, jones}, the transition changes its character from the $2^{\text {nd}}$ order to the $1^{\text {st}}$ one. It was shown by the DSC studies that the ferroelectric phase transition \cite{somaily, chapagain} appearing, in dependence on the composition, within the range $350 - 400$~K is hardly visible in the temperature dependence of specific heat for majority of compositions. We attribute this fact to the wide temperature hysteresis of this transition \cite{somaily, chapagain} and presence of the nanoscale tetragonal distortions in the paraelectric phase \cite{jones}, such that the transition goes through the phase coexistence state over a large temperature interval and thus it is not visible as a maximum of the specific heat.

\begin{acknowledgments}
This work was supported partially by the National Science Centre, Poland, under project No. 2018/31/B/ST5/03024.
\end{acknowledgments}

%\bibliography{Papier-SrBa_MnTi_O3_prb}


\end{document}